\documentclass[twocolumn, trackchanges]{aastex631}
\usepackage{lipsum}% for placeholder text

\usepackage{soul}

\begin{document}
\makeatletter
\let\frontmatter@title@above=\relax
\makeatother

%Define Commands
\newcommand\lsim{\mathrel{\rlap{\lower4pt\hbox{\hskip1pt$\sim$}}
\raise1pt\hbox{$<$}}}
\newcommand\gsim{\mathrel{\rlap{\lower4pt\hbox{\hskip1pt$\sim$}}
\raise1pt\hbox{$>$}}}

%Colors for edits
\newcommand{\CS}[1]{{\color{red}  CS: #1}}
\newcommand{\YH}[1]{{\color{blue} YH: #1}}
\newcommand{\BH}[1]{{\color{blue} BH: #1}}
\newcommand{\TYMY}[1]{{\color{blue} TYMY: #1}}
\newcommand{\RH}[1]{{\color{blue} RH: #1}}

\newcommand{\Kepler}{\textit{Kepler}}
\newcommand{\EPOS}{\texttt{EPOS}}
\newcommand{\norm}[1]{\lvert #1 \rvert}

%\begin{document}

\title{\Large Predicting the Dominant Formation Mechanism of Multi-Planetary Systems}

\author[0000-0003-1247-9349]{Cheyanne Shariat}
\affiliation{Department of Physics and Astronomy, University of California, Los Angeles, CA 90095, USA}
\affiliation{Jet Propulsion Laboratory, California Institute of Technology, Pasadena, CA 91109, USA}

\author[0000-0002-9017-3663]{Yasuhiro Hasegawa}
\affiliation{Jet Propulsion Laboratory, California Institute of Technology, Pasadena, CA 91109, USA}

\author[0000-0001-7840-3502]{Bradley M.S. Hansen}
\affiliation{Department of Physics and Astronomy, University of California, Los Angeles, CA 90095, USA}

\author[0000-0002-5228-7176]{Tze Yeung Mathew Yu}
\affiliation{Department of Physics and Astronomy, University of California, Los Angeles, CA 90095, USA}

\author[0000-0003-2215-8485]{Renyu Hu}
\affiliation{Jet Propulsion Laboratory, California Institute of Technology, Pasadena, CA 91109, USA}
\affiliation{Division of Geological and Planetary Sciences, California Institute of Technology, Pasadena, CA 91125, USA}

\keywords{multi-planetary systems: exoplanets - protoplanetary disks - formation}

\correspondingauthor{Cheyanne Shariat}
\email{cheyanneshariat@ucla.edu}

\begin{abstract}
Most, if not all, sun-like stars host one or more planets, making multi-planetary systems commonplace in our galaxy. We utilize hundreds of multi-planet simulations to explore the origin of such systems, focusing on their orbital architecture. The first set of simulations assumes in-situ assembly of planetary embryos, while the second explores planetary migration. After applying observational biases to the simulations, we compare them to $250+$ observed multi-planetary systems, including $13$ systems with planets in the habitable zone. For all of the systems, we calculate two of the so-called statistical measures: the mass concentration ($S_{\rm c}$) and orbital spacing ($S_{\rm s}$). After analytic and empirical analyses, we find that the measures are related to first-order with a power law: $S_{\rm c} \sim S_{\rm s}^\beta$. The in-situ systems exhibit steeper power-law relations relative to the migration systems. We show that different formation scenarios cover different regions in the $S_{\rm s} - S_{\rm c}$ diagram with some overlap. Furthermore, we discover that observed systems with $S_{\rm s} < 30$ are likely dominated by the migration scenario, while those with $S_{\rm s} \geq 30$ are likely dominated by the in-situ scenario. We apply these criteria to determine that a majority ($62\%$) of observed multi-planetary systems formed via migration, whereas most systems with currently observed habitable planets formed via in-situ assembly. This work provides methods of leveraging the statistical measures ($S_{\rm s}$ and $S_{\rm c}$) to disentangle the formation history of observed multi-planetary systems based on their present-day architectures.

\end{abstract}

\section{Introduction}\label{sec:introduction}

How do planetary systems form? 
This is a fundamental question in astrophysics.
Until recently, the solar system was the only known sample, 
and hence the canonical view of planet formation was developed based on this specific system \citep[e.g.,][]{1981PThPS..70...35H,1984Icar...60..553W}.

The discovery of exoplanets entirely changes the landscape of planet formation.
In particular, NASA’s Kepler mission \citep{Borucki10} generated a surge in observed exoplanets.
We now recognize that, statistically, almost every star hosts a planet, and half of all sun-like stars host a rocky planet in their habitable zone \citep[e.g.,][]{Hsu19, Bryson21}. 
Also, exoplanets exhibit a wide range of architectures that contrast with the configuration of our solar system \citep[e.g., see][for reviews]{Winn15,2021ARA&A..59..291Z, Weiss23}.
These astonishing discoveries challenge the canonical view of planet formation.

One of the most exciting discoveries is an abundant population of a new class of exoplanets -- close-in small-sized planets.
Currently, at least two formation scenarios are actively investigated in the literature: the in-situ and migration scenarios.
The in-situ scenario proposes that planets formed near where they are observed today\citep[e.g.,][]{2010ApJ...719..810I,Hansen12,Chiang13}. This scenario is characterized by large, protoplanetary embryos that undergo at least one giant impact.
Importantly, in-situ formation is unlikely to explain the formation of all multi-planetary systems \citep{Raymond14, He22}.

The competing formation model is the migration scenario.
In this scenario, once (proto)planets have become massive enough in the protoplanetary disk, they excite spiral density waves that provide a gravitational torque, leading to migration of the planets \citep{Lin79, Goldreich80, Ward97}. 
Planets may lose angular momentum until they spiral into the inner edge of the disk, where the positive co-rotation torque halts them \citep[e.g.,][]{Masset06, Kley12}. 
One robust prediction of the migration scenario is that planetary systems are captured predominantly by mean motion resonances \citep[MMRs, e.g.,][]{Terquem07,Hellary12,Izidoro17, Ogihara18}.
The period ratio distribution of observed exoplanetary systems, however, does not support this prediction \citep[e.g.,][]{Winn15,Batygin15}, and hence the migration scenario alone cannot reproduce the Kepler observations, either.
Therefore, previous studies imply that simultaneous examination of various formation scenarios would be critical to fully reproduce the observed population of exoplanets.
%\

Here, we show that combining both the in-situ and migration scenarios can lead to a better understanding of the origin of observed multi-planet systems.
We provide analytic predictions to contrast characteristics of the competing formation mechanisms (Section \ref{sec:analytical_prediction}).
We also outline methods of determining the dominant formation mechanism of an observed planetary system based on its present-day architecture, including habitable planetary systems (Section \ref{sec:Determination_formation}).
This work thus demonstrates that a huge diversity of multi-planet systems can be governed by two dominant formation scenarios: in-situ and migration.

\section{Analytical prediction}\label{sec:analytical_prediction}

We first provide our analytical predictions, 
wherein two competing formation scenarios (in-situ vs migration) are compared.
To proceed, we adopt the so-called statistical measures (SMs) -- 
quantities that characterize different aspects of a multi-planetary system \citep{Chambers01}.
Here we discuss how the two formation scenarios can be differentiated by SMs.

\subsection{Characteristic Separations}\label{subsec:characteristic_separations}

Different formation scenarios may end up with different separations between nearby planets.
Two characteristic separations can be defined in this work: $\Delta_{\rm in-situ}$ and $\Delta_{\rm mig}$.
The former involves the in-situ scenario, while the latter focuses on the planetary migration scenario.

When multi-planet systems are formed by the in-situ scenario,
$\Delta_{\rm in-situ}$ should be determined by 
the result of the pure gravitational interaction between neighboring protoplanets.
This separation is achieved before they undergo a giant impact,
that is, resultingly formed planetary systems should have the mutual separation of $> \Delta_{\rm in-situ}$.
Mathematically, it can be written as \citep[e.g.,][]{Schlichting2014}
\begin{equation}
\Delta_{\rm in-situ} \simeq 2 a_{\rm p} \frac{v_{\rm esc}}{v_{\rm Kep}},
\end{equation}
where $a_{\rm p}$ is the semimajor axis of (proto)planets, $v_{\rm Kep}$ is the Keplerian velocity, and $v_{\rm esc}= \sqrt{2GM_{\rm p}/R_{\rm p}}$ is the escape velocity for the neighboring (proto)planets with comparable masses ($M_{\rm p}$) and radii ($r_{\rm p}$).

Thus, the characteristic separation between two planets set by the in-situ scenario is given as
\begin{eqnarray}
\label{eq:delta_in-situ}  
\frac{ \Delta_{\rm in-situ} }{a_{\rm p}} & \simeq & 2 \frac{v_{\rm esc}}{v_{\rm Kep}}  \\ \nonumber
                                              & \simeq & 0.34 \left( \frac{M_{\rm p}}{ 2 M_{\oplus} } \right)^{1/2} \left( \frac{R_{\rm p}}{ 1.5 R_{\oplus} } \right)^{-1/2} \\ \nonumber
                                              & \times  & \left( \frac{M_{\rm s}}{ 1 M_{\odot} } \right)^{-1/2} \left( \frac{a_{\rm p}}{ 0.3 \mbox{ au} } \right)^{1/2},                                            
\end{eqnarray}
where $M_{\rm s}$ is the mass of the host star.
This indicates that if planetary systems are formed by the in-situ scenario, 
then the spacing should be larger than $\sim 0.3$.

On the other hand, if planetary migration plays an important role in forming multi-planet systems,
then the characteristic separation should be computed by two competing processes \citep{Ida10, Arora21}:
Convergent migration, which reduces the mutual spacing between neighboring protoplanets,
is compensated by the gravitational repulsion between them.
The common outcome is that nearby planets are captured in MMRs.
Many numerical simulations show that migrating planets are captured in the 2:1 period resonances \citep[e.g.,][]{Izidoro17,Yu23}.
Then, $\Delta_{\rm mig}$ can be computed using Kepler's Third Law as
\begin{equation}
\left( \frac{ a_{\rm p} + \Delta_{\rm mig} }{ a_{\rm p} } \right)^{3/2} = 2.
\end{equation}
Equivalently,
\begin{equation}
 \frac{ \Delta_{\rm mig} }{ a_{\rm p} } = 2^{2/3}-1 \simeq 0.59.
 \label{eq:delta_mig}
\end{equation}
This suggests that if convergent planetary migration defines the characteristic separation,
then it should be smaller than $\sim 0.6$.

\subsection{The Statistical Measures}\label{subsec:statistical_measures}

Characterizing multi-planet systems is not straightforward;
as the number of planets in a system increases, 
so does the number of quantities needed to fully characterize the system. 
\citet{Chambers01} devised a series of SMs to compare the architecture of different multi-planetary systems. 

We focus on two of these quantities as done in previous studies \citep[e.g.,][]{Sun17, Clement19, Lykawka19, Arora21, Mah22}.  
The first quantity is a measure of the radial mass concentration ($S_{\rm c}$) and is given by 
\begin{equation}\label{eq:Sc}
S_{\rm c} = \mbox{ max} \left( \frac{\sum_{j} M_{{\rm p},j}}{{\sum_j M_{{\rm p},j}[\log_{10}(a/a_{{\rm p},j})]^2}} \right),
\end{equation}
where $M_{{\rm p},j}$ ($a_{{\rm p},j}$) is the mass (semi-major axis) of the $j^{\text{th}}$ planet from the host star, and $a$ is a variable denoting the distance from the host planet.
A certain value of $a$ is chosen such that $S_{\rm c}$ takes the maximum value.
Compact planetary systems with similar masses and low multiplicities tend to have a high value of mass concentration. 
The value of $S_{\rm c}$ is essentially a measure of how evenly spaced the planet's masses are in the system.

The second quantity that we employ is the orbital spacing ($S_{\rm s}$), 
which is similar to the averaged mutual spacing normalized by the mutual Hill radius.
One difference is that $S_{\rm s}$ is normalized by $M_{\rm p}^{1/4}$ (not $M_{\rm p}^{1/3}$). 
This is motivated by the results of $N$-body simulations that explore the stability of multi-planet systems \citep{Chambers96}:
\begin{equation}\label{eq:Ss}
S_{\rm s} = \frac{6}{N-1} \left( \frac{a_{\rm p,max} - a_{\rm p,min}}{a_{\rm p,max} + a_{\rm p,min}}\right) \left( \frac{3 M_{\rm s}}{2 \overline{M}_{\rm p}}\right)^{1/4},
\end{equation}
where $N$ is the number of planets in the system, $a_{\rm p,max}$ ($a_{\rm p,min}$) is the maximum (minimum) semi-major axis, and $\overline{M}_{\rm p} (= \sum_j M_{{\rm p},j} /N)$ is the mean mass of planets in the system.

% and the angular momentum deficit ($S_d$)
% \begin{equation}
% S_d = \sum_{j=1}^N \frac{M_{p,j}}{M_\star}\sqrt{a_j} \left( 1 - \sqrt{1-e_j^2}~cos~i_{m,j}\right). 
% \end{equation}

\subsection{The Two Planet Case}\label{subsubsec:two_planets}

We here consider the two-planet system case and examine how the SMs can be used to differentiate the two formation scenarios.

Our mathematical manipulation leads to a simplified expression of mass concentration  (Appendix \ref{app:SMs_2planets}), which is given as
\begin{equation}
    S_{\rm c} =  \frac{ ( M_{{\rm p},1} + M_{{\rm p},2} )^2 }{ M_{{\rm p},1} M_{{\rm p},2} }  \frac{1}{[ \log_{10}{(a_{{\rm p},2}/a_{{\rm p},1})}]^2}.  
\end{equation}
For orbital spacing, 
\begin{equation}
    S_{\rm s} = 6 \Bigg( \frac{a_{{\rm p},2} - a_{{\rm p},1}}{a_{{\rm p},2}+a_{{\rm p},1}}\Bigg)\Bigg(\frac{3 M_{\rm s}}{ M_{{\rm p},1} + M_{{\rm p},2} }\Bigg)^{1/4}.  
\end{equation}

Under the assumptions that $M_{{\rm p},1} \simeq M_{{\rm p},2}$ and that $\Delta a / a_{{\rm p},1} \ll 1$, where $a_{{\rm p},2} = a_{{\rm p},1} + \Delta a$,
the above two equations can be expressed to first order as
\begin{equation}
    S_{\rm c} \simeq  4  \left( \frac{\ln(10) }{ \Delta a / a_{{\rm p},1} } \right)^2 \simeq 2.4 \times 10^2   \left( \frac{ \Delta a / a_{{\rm p},1} }{0.3 } \right)^{-2},
    \label{eq:Sc_two}
\end{equation}
and 
\begin{eqnarray}
\label{eq:Ss_two}
    S_{\rm s}  & \simeq & 3 \Bigg(\frac{3 M_{\rm s}}{ 2 M_{{\rm p},1} }\Bigg)^{1/4}  \Delta a / a_{{\rm p},1}  \\ \nonumber
            & \simeq & 2.0 \times 10 \left( \frac{ M_{{\rm p},1} }{ 2 M_{\oplus} } \right)^{-1/4}  \left( \frac{ M_{\rm s} }{ 1 M_{\odot} } \right)^{1/4} \left( \frac{ \Delta a / a_{{\rm p},1} }{0.3} \right).
\end{eqnarray}

We finally combine the above two equations and find that
\begin{equation}
\label{eq:S_cS_s}
    S_{\rm c} \simeq   4.7 \times 10^3  \left( \frac{ M_{{\rm p},1} }{ 2 M_{\oplus} } \right)^{-1/4}  \left( \frac{ M_{\rm s} }{ 1 M_{\odot} } \right)^{1/4}  S_{\rm s}^{-2}.
\end{equation}

Equation (\ref{eq:S_cS_s}) suggests that similar mass planets that are relatively in tight orbits will exhibit an inverse-square relationship between the SMs. This is a crucial prediction and will be juxtaposed with the simulation results in later sections.

\subsection{Predictions}\label{subsec:predictions}

We now apply the above-simplified equations (i.e., equations (\ref{eq:Sc_two}) and (\ref{eq:Ss_two})) to the in-situ and migration scenarios to examine whether these two scenarios can be differentiated.

As discussed in Section \ref{subsec:characteristic_separations}, 
if planetary systems are formed by the in-situ scenario,
then $\Delta_{\rm in-situ} / a_{\rm p} \ga 0.3$ (equation (\ref{eq:delta_in-situ})).
This leads to $S_{\rm c} \la  3 \times10^2$ and $S_{\rm s} \ga 20$.

On the other hand, if planetary systems are formed by the migration scenario,
then $\Delta_{\rm mig} / a_{\rm p} \la 0.6$ (equation (\ref{eq:delta_mig})).
This leads to that $S_{\rm c} \ga 60$ and $S_{\rm s} \la 40$.

Thus, these calculations suggest that dominant formation mechanisms could be identified by computing mass concentration and orbital spacing,
and switching a formation mechanism could occur at the ranges of $60 \la S_{\rm c} \la 300$ and  $20 \la S_{\rm s} \la 40$.

These predicted ranges will be compared to the results of our simulations in Section \ref{subsec:testing_analytical_predictions}. Despite being derived from the only two-planet case, the inequalities are robust and can be generalized to systems with three and four planets as well.

\section{Determination of a formation scenario} \label{sec:Determination_formation}

We test the above predictions, using both simulation results and observational data available in the literature.
Since exoplanetary systems observed by Kepler suffer from various observational biases,
one cannot compare simulation results with observational data directly.
We therefore apply observational biases to simulated planetary systems to resolve this issue.
%For brevity, we hereafter refer to such biased, simulated planetary systems as synthetic systems,
%while exoplanetary systems observed by Kepler as observed systems.
We use two kinds of simulations that study the in-situ and migration scenarios.
We compare simulation results with observational data to identify a formation scenario of observed systems.

\subsection{Observed Systems}\label{subsec:obs_sample}
We first describe observed systems.
We gather a sample of multi-planetary systems from the catalog of the Kepler Space Telescope, 
following the filtering methods described in \citet{Arora21}. 
This observed sample is taken from \citet{keplercumulativetable}\footnote{
The source data was obtained from https://exoplanetarchive.ipac.caltech.edu/cgi-bin/TblView/nph-tblView?app=ExoTbls\&config=cumulative
on 2021-01-21 at 16:17 with the size of 9564 $\times$ 49 columns.} 
and contains systems with at least two planets orbiting around a main sequence star. 
After the filtering, we are left with $234$ observed systems ($837$ planets total) with two or more planets in each system.

\subsection{Simulation Data}\label{subsec:simulation_setup}

We use the data from two sets of simulations. 
The first set contains 100 realizations of `in-situ' simulations from \citet{Hansen13, Hansen15}. These models assume that the protoplanetary disk has a surface density profile of $\Sigma \sim a^{-1.5}$, where $a$ is the distance from the $1$~M$_\odot$ host star. The surface profile of the disk is then normalized to contain $20$~M$_\oplus$ of rocky material interior to $a=1$~au. The simulations are run for $10$~Myr in $12$~hr time steps.
The in-situ model assumes in-place planetary embryos and giant impact(s) to be dominant in forming multi-planetary systems.

The second set contains 538 planetary systems that undergo two-planet migration (Yu et al. 2024, in preparation). These `migration' simulations explore the effects of two-planet migration on the final architecture of systems around a $1$~M$_\odot$ host star. The disk structure and simulation procedure are outlined in \citet{Yu23}. These simulations assume that the planets initially take on integer values between $1$ and $10$ M$_\oplus$, and initial semi-major axes between 0.3 and 1.2 AU. Initial eccentricities and inclinations are both set to 0.1, and the initial period ratio is chosen to be $2.6$ between the two planets to avoid artificial capture in a mean motion resonance.
The initial orbital configuration is not crucial as it is washed out by migration.

For more details on the in-situ and migration simulations, refer to \citet{Hansen13} and \citet{Yu23}, respectively.

We note that these simulations did not provide planet radius, so we employ the probabilistic algorithm from \citet{CK17} to calculate the radius.

\subsection{Biasing Simulation Data}\label{subsec:biasing_sim_data}

To meaningfully compare the simulation results to observations from Kepler, we bias the simulation data. 
We employ the Exoplanet Population Observation Simulator \citep[$\EPOS$;][]{EPOSI,EPOSII} to produce simulated observations of the theoretical systems. 
$\EPOS$ generates simulated Kepler observations of our multi-planetary systems, using the Kepler DR25 final catalog \citep{Coughlin17}. The software utilizes planetary parameters (masses, semi-major axes, eccentricities, inclinations) as inputs and adopts a Monte Carlo approach to sample the geometry of the systems. From these simulated detections, $\EPOS$ supplies the various combinations of planets that could be recovered by Kepler, with the respective probabilities for each system of planets.

\begin{figure*}
\includegraphics[width=1.0
\textwidth]{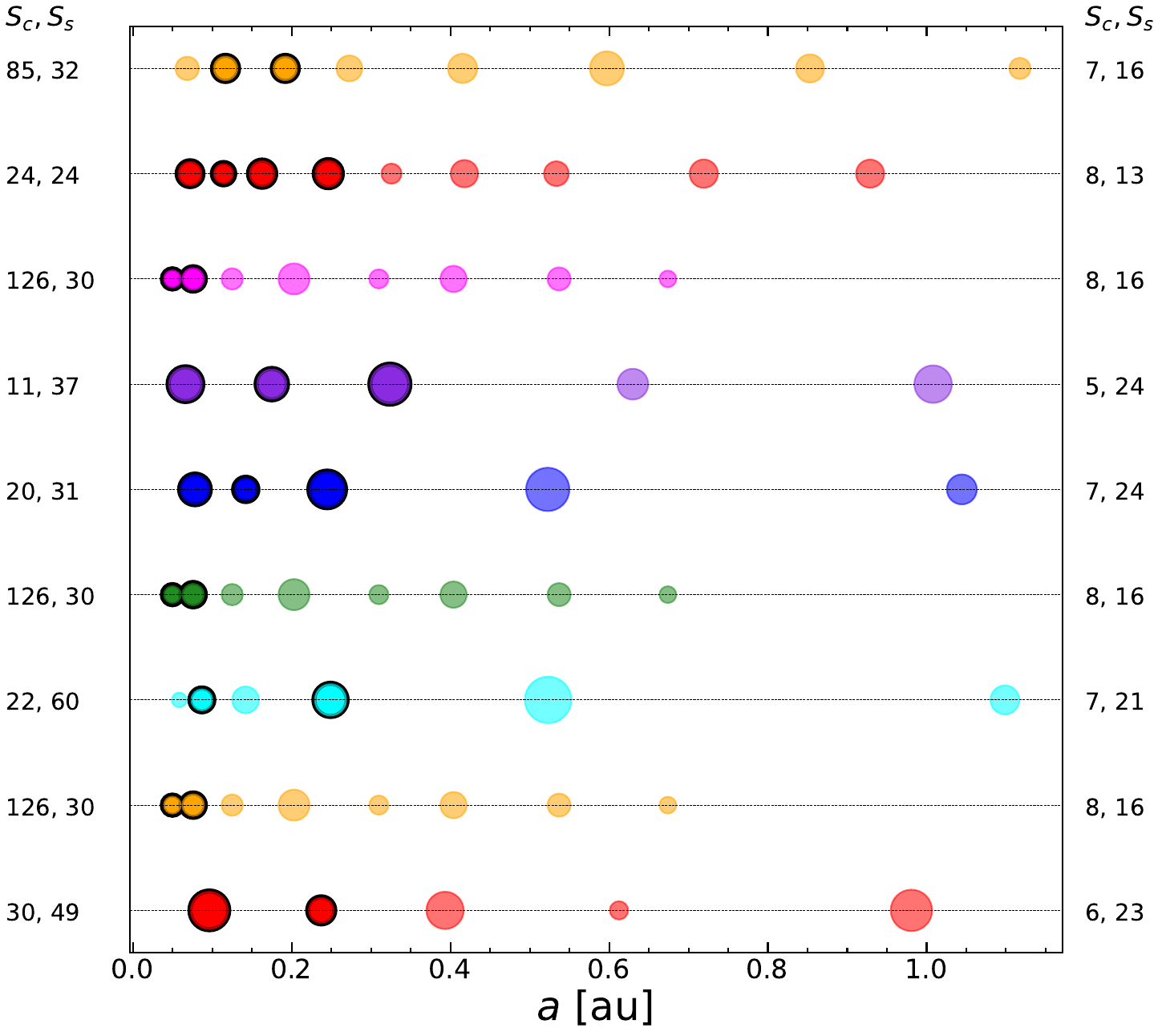}

\figcaption{The planetary architectures of the simulated in-situ multi-planetary systems before and after biasing. The planets outlined in black represent the observed planets after the simulations were biased using \texttt{EPOS}. All planets in a given row represent the architecture of the inherent system as a result of the in-situ simulations. The sizes of the planets are scaled by the mass, and the horizontal axis represents their distance from the host star in au. We write the mass concentration ($S_{\rm c}$) and orbital spacing ($S_{\rm s}$) of the system after it was biased for each system on the left, and those of the inherent system on the right. Note that the statistical measures change, often significantly, after observational bases are applied to the simulations. 
\label{fig:bias_arch}} 
\end{figure*}

We plot the architecture of a subset of the in-situ simulated systems before and after being biased using $\EPOS$ in Figure \ref{fig:bias_arch}. Each row represents a multi-planetary system with the semi-major axis on the horizontal axis, and the size of each planet is scaled by its mass. The outlined planets are the ones that were detected from our simulated Kepler observations using $\EPOS$ (i.e., the `biased systems'). 
As expected, the more massive (i.e., larger radii) and short-period planets are recovered from the simulated observations. 

We also compute the values of the SMs ($S_{\rm c}$, $S_{\rm s}$) for both biased and inherent systems and label them on the left and right sides of Figure \ref{fig:bias_arch}, respectively.
The SMs of a given system often change significantly after biasing. 
In most cases, both the mass concentration ($S_{\rm c}$) and orbital spacing ($S_{\rm s}$) increase, leaving little (or no apparent) trace of the hidden, more distant planets. Particularly, biased planetary systems tend to have more concentrated mass and wider spacing than what is truly present. Additional discussions about how $\EPOS$ biases our simulated systems are outlined in Appendix \ref{app:EPOS}.

After biasing, we find that there are $536$ migration systems and $88$ in-situ systems with $\geq2$ planets that are detectable by Kepler. For the in-situ samples, biasing the data greatly decreased the multiplicity of the systems. This occurs because missed planets are less massive with non-zero inclination and/or long orbital periods. On the other hand, the migration simulations originally consisted of systems with two massive, short-period planets, and hence the biasing from $\EPOS$ only removed a few systems. For systems with three or more planets, the inner planet can also become caught in a secular resonance, especially when the innermost planet is less massive than its companions. \citet{Faridani23} show that these secular resonances can serve to increase the eccentricity and mutual inclination of this planet over short timescales, which would also impact the detectability of such planets. $\EPOS$ does not consider this effect when biasing.

\subsection{Testing Analytical Predictions}\label{subsec:testing_analytical_predictions}

We test our analytical predictions against simulation data.

\begin{figure*}
\centering
\includegraphics[width=0.95
\textwidth]{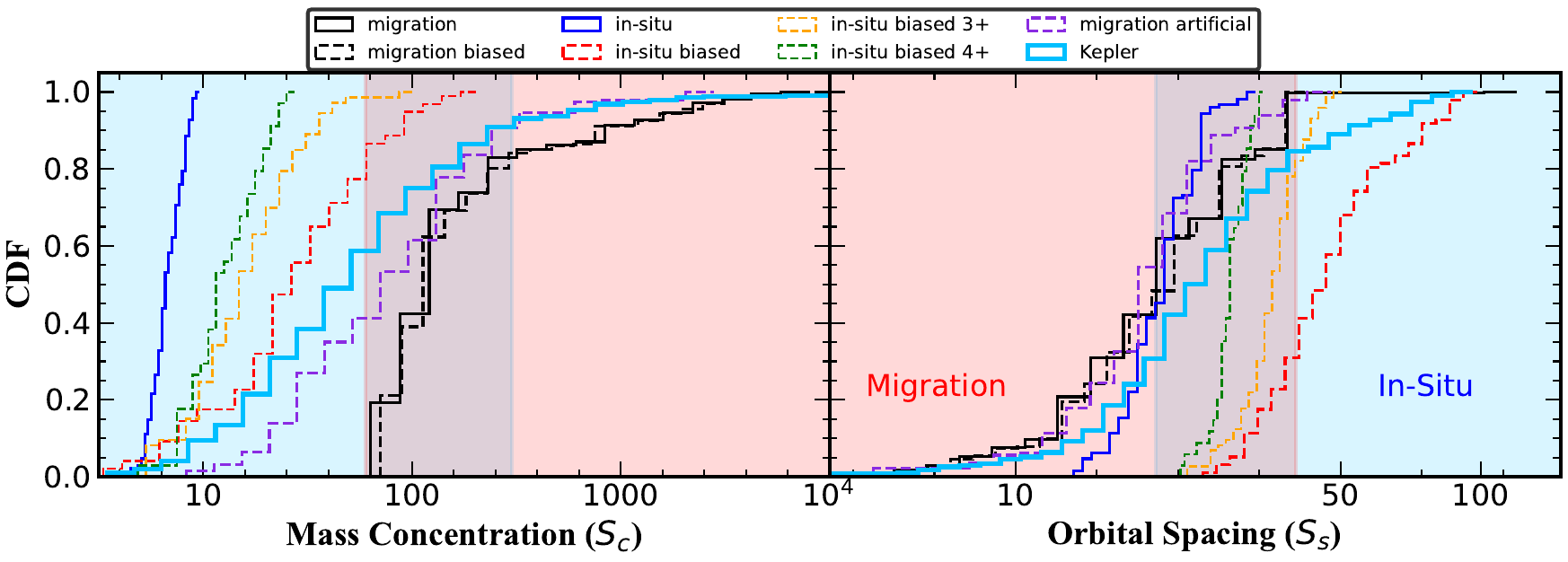}
\figcaption{
The cumulative distribution functions (CDFs) of the \citet{Chambers01} statistical measures for `in-situ' and `migration' simulations, before and after biasing the data. %, compared to the observed catalog. 
The raw migration data (black) and raw in-situ data (blue) are compared to the biased migration data (dashed black) and the biased in-situ data (dashed red). 
 For the biased data, we require that the number of planets observed in the system is at least two, making it a multi-planetary system. 
For the in-situ data, %we also plot the SMs of the system containing only the inner two planets (dashed blue). Moreover, we look to separate the in-situ biased systems by the minimum number of planets that are observed. We 
we split it by the minimum number of observed planets, namely, %the minimum observed of 
3 and 4 planets, which are shown in orange and green, respectively.
The red (blue) shaded regions correspond to the ranges derived from analytic methods in Section \ref{subsubsec:two_planets} and are consistent with the migration (in-situ) formation scenario.
The artificial migration data is plotted by the dashed purple line, which shows some deviation from our analytical prediction at low $S_{\rm c}$, while such deviation is not seen for $S_{\rm s}$.
For comparison purposes, the SMs of the Kepler data are also plotted (the thick light blue).
\label{fig:SMs} }

\end{figure*}

In Figure \ref{fig:SMs}, we display the cumulative distribution of the SMs for the simulation data before and after biasing. 
For the in-situ simulations,
we explicitly denote biased systems that have $\geq3$ and $\geq4$ planets by the orange and green dashed curves, respectively. The red curve displays the full set of `in-situ biased' systems, which has no restrictions on the multiplicity of the biased systems.

Because the migration simulations cover only the two-planet case, we also generate an artificial set of migration systems with higher multiplicities. To pursue this, we sample the multiplicities of the artificial systems to match the multiplicity distribution of the biased in-situ samples, which represent the observed systems well.
Then, for a chosen multiplicity, we 
assign the semi-major axis of the innermost planet
from the distribution of 
the simulated migration systems.
The semi-major axis of the $j^{\text{th}}$ planet is computed recursively, starting with the innermost planet. We sample a semi-major axis ratio value from the distribution of the migration simulations and multiply it with the semi-major axis of the $(j-1)^{\text{th}}$ planet to generate the semi-major axis of the $j^{\text{th}}$ planet, mimicking the capture of a MMR.

The masses for each planet are chosen to be uniform in the range $3$-$10$~$M_\oplus$, similar to \citet{Yu23}. We combine $500$ of these artificial samples with the $2$-planet migration simulations to get a combined theoretical sample, which we plot with the dashed purple curve in Figure \ref{fig:SMs}. After combining the two-planet migration simulations with the artificial migration sample, we find that the multiplicity distribution is manifestly similar to that of the biased in-situ systems.

%The \YH{resulting} ranges of $S_{\rm s}$ and $S_{\rm c}$ that the \YH{simulated} planets span are different between the simulations. %However, we 
%We confirm that our theoretical predictions are broadly consistent with the simulation results. 
In Figure \ref{fig:SMs}, we also shade the ranges of $S_{\rm c}$ and $S_{\rm s}$ as predicted from our analytical consideration (Section \ref{subsec:predictions}) and find that they are
consistent with the various simulation results, %, finding 
with a small region of overlap in the middle of the distribution. 
We confirm that the migration simulations of the two-planet case generally captured the trend of the higher-multiplicity systems.
We further discuss and constrain the ranges of $S_{\rm s}$ and $S_{\rm c}$ for each formation mechanism %simulation 
in Section \ref{subsec:identifying_formation_mechanism}. 

% To investigate this relation further, we calculate the SMs for a two-planet system, assuming the planets to be close to one another (see Appendix \ref{app:multiplicity_SMs} for calculation). 

\begin{figure*}
\includegraphics[width=0.95
\textwidth]{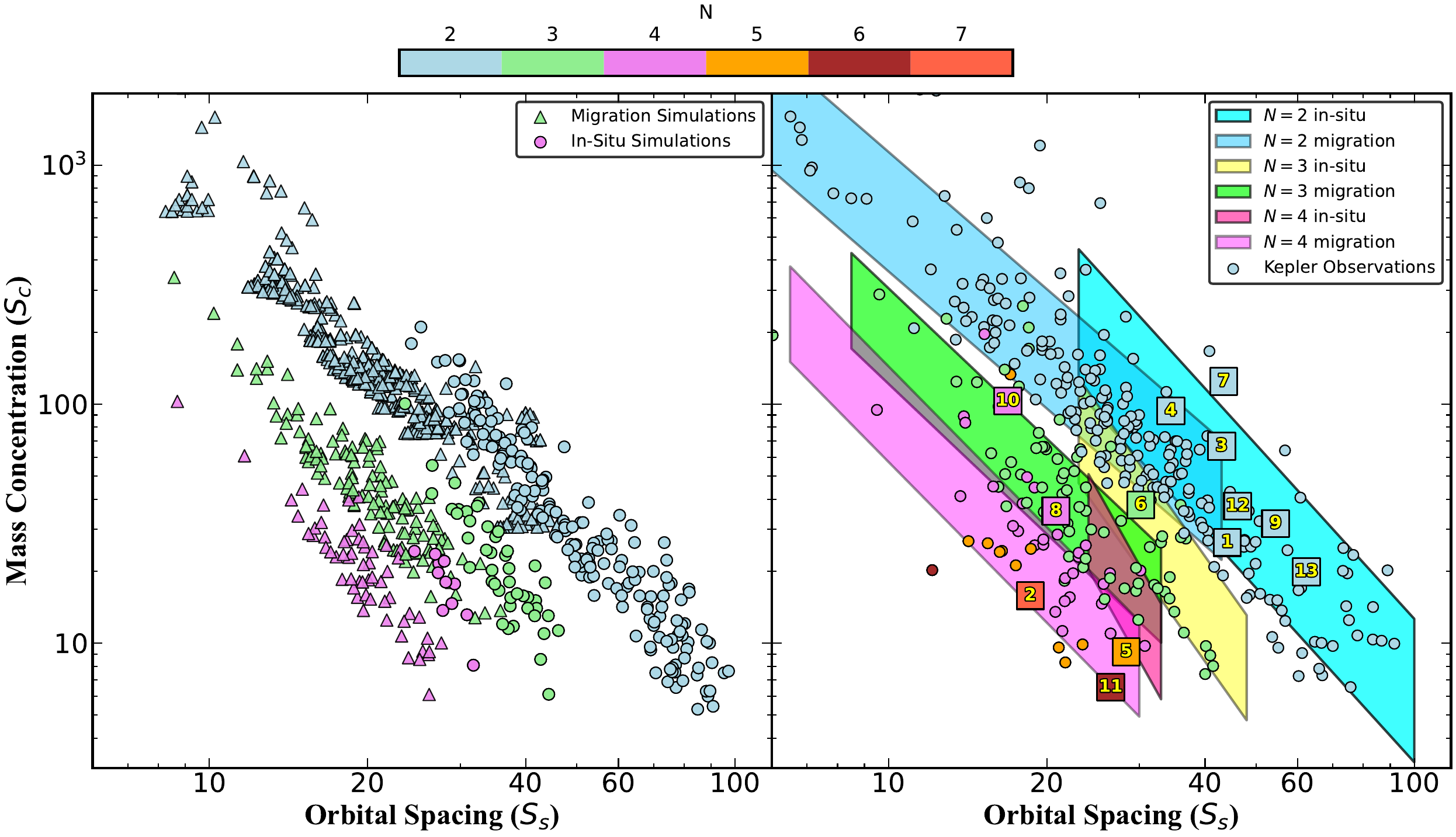}
\figcaption{
The mass concentration as a function of the orbital spacing for the simulated in-situ (circle) and migration (triangle) systems (left) as compared to observed Kepler systems (right).
%observed Kepler in-situ and Kepler migration systems, as compared to the simulated in-situ (circle) and migration (x-shaped) systems. 
The colorbar represents the multiplicity of the system. 
On the left, biased simulation data shows a power-law relationship between $S_{\rm c}$ and $S_{\rm s}$, 
and its slope is likely to be a function of multiplicity, especially for the in-situ systems.
On the right, the distributions of biased simulation data are denoted by the shaded regions.
The Kepler observed systems are shown as the circles.
The numbered squares display the statistical measures for thirteen observed multi-planetary systems that host at least one (or more) planet(s) inside the habitable zone. They are (1) GJ 667C, (2) Trappist 1, (3) Teegarden's Star, (4) GJ 1002, (5) Kepler-186, (6) GJ 1061, (7) Proxima Centauri, (8) K2-72, (9) GJ 273, (10) TOI-700, (11) Kepler-62, (12) LP 890-9, and (13) LHS 1140. See Appendix \ref{app:hab_systems} for further details on these habitable planetary systems.
\label{fig:kep_MGIS_SM_scatter} }

\end{figure*}

We finally test our predictions by examining the relationship between $S_{\rm c}$ and $S_{\rm s}$.
In Figure \ref{fig:kep_MGIS_SM_scatter}, we plot $S_{\rm c}$ as a function of $S_{\rm s}$ for all of our biased simulation data (left panel). We color the scatter points based on their multiplicities and discover many interesting features. Firstly, we find that, for both simulations, the populations with the same multiplicity are linear in a log-log scale. This suggests a power-law relationship between the statistical measures: 
\begin{equation}
    S_{\rm s} \sim (S_{\rm c})^\beta.
\end{equation}

%In Section \ref{subsubsec:two_planets} we have predicted the relationship that $S_{\rm s} \sim S_{\rm c}^{-2}$ for the two-planet case, under the assumption that the planets are tight ($\Delta a / a_{p,1} \ll 1$). Since the migration simulations produce such close systems, we would predict this slope to be present in the migration samples with  $N=2$ planets. We indeed find that this is consistent with the simulations.

Secondly, %After performing a least-squares fit, 
we determine the best-fit value of $\beta$ for each of the subpopulations by performing a least-squares fit. %, which are denoted with different markers and colors in Figure \ref{fig:kep_MGIS_{\rm s}M_scatter}. 
For the simulated migration systems (squares in the left panel of Figure \ref{fig:kep_MGIS_SM_scatter}), we find that the best-fit power-law slopes are $\beta = -2.03, -2.20, -2.51$ for systems with multiplicities of $N=2, 3, 4$, respectively.
For the simulated in-situ systems (circles in the left panel of Figure \ref{fig:kep_MGIS_SM_scatter}), we find that the best-fit power-law slopes are $\beta = -2.42, -3.11, -3.93$ for systems with multiplicities of $N=2, 3, 4$, respectively. Notably, the $N=2$ migration systems exhibit $\beta \approx 2$, as predicted in Equation (\ref{eq:S_cS_s}). A similar slope is observed in the $N=3$ migration systems, which suggests that the close-spacing assumption is a key characteristic of migration systems. %More generally, 
Finally, we find that, within a multiplicity category, the in-situ systems exhibit steeper power-law slopes than the migration systems. 

In the right panel of Figure \ref{fig:kep_MGIS_SM_scatter}, 
%we plot the SMs for our observed sample of multi-planetary systems (circles), along with 12 known planetary systems with habitable planets (stars). 
we define the ranges of the parameter space that are occupied by the various simulation populations. 
The ranges are chosen by plotting a minimum and maximum power law profile, based on the $\beta$ values above, for a given multiplicity. These lines are then set to cover the minimum and maximum SM ranges reached by the simulations (scatter points in the left panel of Figure \ref{fig:kep_MGIS_SM_scatter}). After this procedure, we can interpret the ranges of $S_{\rm s}-S_{\rm c}$ space that are occupied by the migration and in-situ simulations at multiplicities $2$, $3$, and $4$.
%In particular, we overplot the best-fit line spread across the range occupied by the simulation data on top of the observations. 

In the following section, we leverage the ranges to specify 
%right panel to predict 
the dominant formation mechanism for each observed system. 
%For the region with overlapping theoretical curves, we can probabilistically assign a dominant formation mechanism for the observed systems by analyzing at the specific value of the SMs. 
%In the following section, we will analyze and outline the ranges of SMs exhibited by the different formation mechanisms.

% We note that the migration simulations only test two-planet migration, it is common to observe chains of planets with each pair exhibiting a mean motion resonance \citep[see][for a review]{Batygin15}. To determine the impact of higher multiplicity resonant (migration) systems, we add 100 new systems into our migration sample with multiplicities of $3$, $4$, and $5$, where all of the planets are in a random resonant chain configuration (either 2:1 or 3:2). In all, we find that, although it produces some systems with smaller orbital spacings and mass concentrations, the overall distributions do not change significantly. In Appendix \ref{app:MMR}, we further outline our method of sampling new systems and illustrate the effect on the SM distributions. From this, we conclude that the dominant dynamical phenomenon that contributes to distinguished migration from in-situ systems is the presence of mean-motion resonances. Therefore, using the two-planet migration simulations is sufficient for testing the migration formation scenario and discriminating these SMs from the in-situ distribution.

\subsection{Identifying a Formation Mechanism}\label{subsec:identifying_formation_mechanism}

%We aim to classify the observed multi-planetary systems from our Kepler catalog as having a formation history dominated by either in-situ or migration formation. To do this, we perform a deep investigation on the SMs of the Kepler sample, and their relation to our simulation data. 

It is of primary importance that the Kepler distribution is somewhere in the middle of the biased in-situ and migration systems for both the $S_{\rm c}$ and $S_{\rm s}$ curves (Figure \ref{fig:SMs}). This suggests that many multi-planetary systems in the observed sample can be explained by both formation histories. 
We aim to split the Kepler sample into two parts and identify the in-situ- and migration-dominated systems, based on mass concentration and orbital spacing cutoffs.

%From the SM distribution of the entire Kepler sample (shown in light blue in Figure \ref{fig:SMs}), we aim to separate the in-situ- and migration-dominated systems. We aim to do this by splitting the Kepler sample into two parts, based on the mass concentration and orbital spacing cutoffs. 

In doing so, we are guided by our theoretical predictions outlined in Section \ref{subsec:predictions}. 
%In Section \ref{subsec:predictions}, we asserted that $S_{\rm c} \ga 60$ and $S_{\rm s} \la 40$ is characteristic of a migration-dominated formation history and that $S_{\rm c} \la 300$ and $S_{\rm s} \ga 20$ is characteristic of an in-situ formation history. Although we do not restrict our search to these values, we find 
We will show below that the true cutoff values that split the Kepler systems %accordingly 
lie within the predicted ranges.
In the following, we focus on $S_{\rm s}$ and show that it is a robust signature of the formation mechanism. We also consider a cutoff based on $S_{\rm c}$ in Appendix \ref{subsubsec:split_by_sc}. Additional analyses, including machine learning classification, are summarized in Appendix \ref{app:class}.

%\subsubsection{Splitting by $S_{\rm s}$}\label{subsubsec:split_by_ss}

We test the impact of splitting the observed systems based on an $S_{\rm s}$ criterion. 
%Since systems dominated by a migration formation history are often more tight \citep[e.g.,][]{Goldreich80} than in-situ systems, we expect migration systems to have smaller $S_{\rm c}$ values. 
To find the best orbital spacing cutoff value ($S_{\text{s, cutoff} }$), we consider values ranging from $S_{\text{s, cutoff} } = 10$ to $S_{\text{s, cutoff} } = 40$, motivated by the distribution in Figure \ref{fig:SMs}, and find which values provide the greatest consistency between the SM cumulative distribution functions (CDFs).

\begin{figure*}
\includegraphics[width=0.95
\textwidth]{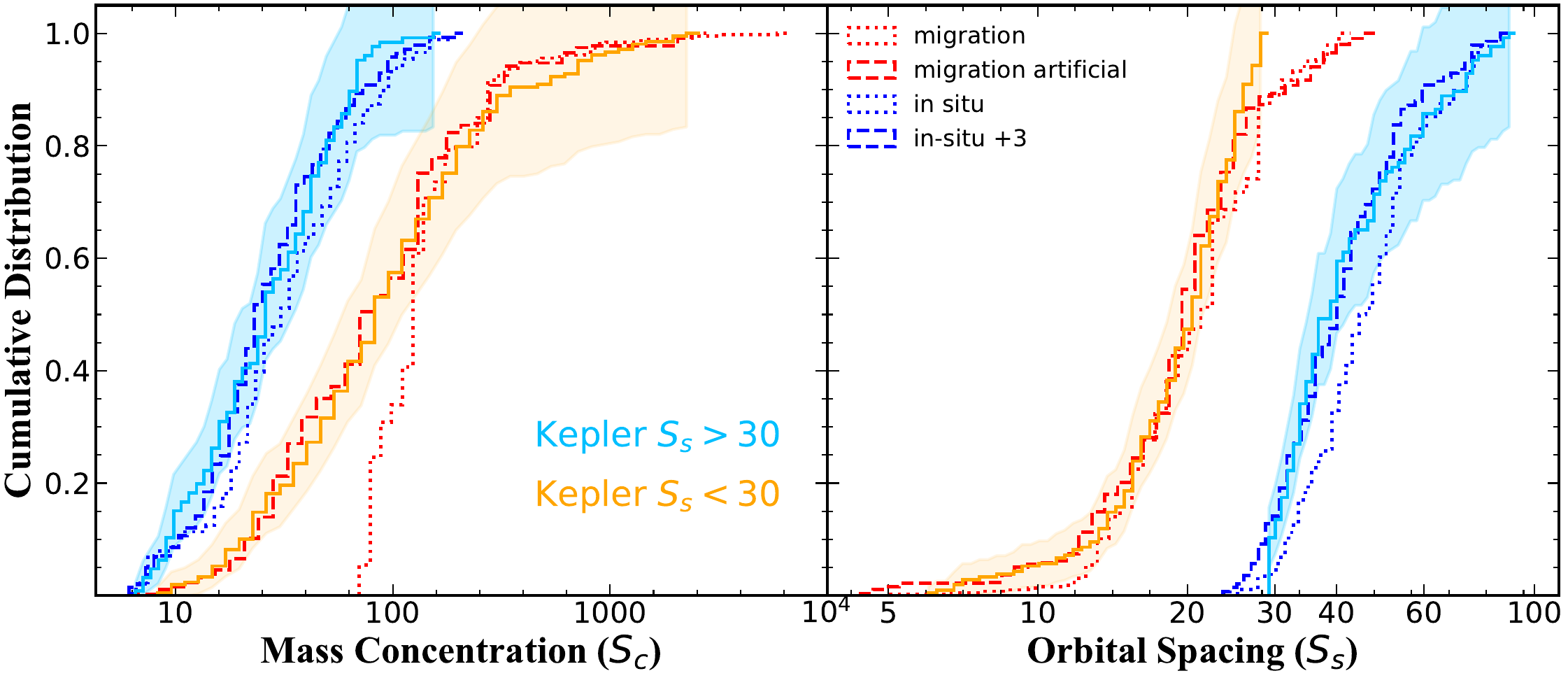}
\figcaption{
The cumulative distributions of the mass concentration ($S_{\rm c}$; left column) and orbital spacing ($S_{\rm s}$; right column) of the biased simulation data compared to %of 
the observed Kepler sample, which are both split into two subgroups.
The dotted blue lines are the biased in-situ simulated systems, and the dashed blue lines are in-situ systems selected for more systems with multiplicities greater than 3. The dotted red lines are the distributions of the biased migration simulation data, and the dashed red lines are the artificial migration systems generated to include higher multiplicities (see Section \ref{subsec:testing_analytical_predictions} for details). We compare this to the observed Kepler systems with different cutoffs. %In the top row, we 
We split the observed systems based on those with $S_{\rm s}<30$ (orange) and $S_{\rm s}>30$ (light blue). %In the bottom row, we split the observed systems based on those with $S_{\rm c}>70$ (orange) and $S_{\rm c}<70$ (light blue). 
For the observed distributions, we include shaded regions that represent Poisson error bars for each bin. 
\label{fig:SMs_ss} }

\end{figure*}

We quantitatively determine $S_{\text{s, cutoff} }$ by selecting the cutoff value that maximizes the sum of the Kolmogorov–Smirnov (KS) test $p$-values, which would suggest that the observed and simulated distributions are unlikely to come from different parent distributions. Here, we split the Kepler sample into systems that have an $S_{\rm s}<S_{\text{s, cutoff} }$, and those with $S_{\rm s}>S_{\text{s, cutoff} }$. Then, we calculate the KS $p$-value between the lower Kepler sample ($S_{\rm s}<S_{\text{s, cutoff} }$) and the biased migration systems, and between the higher Kepler sample ($S_{\rm s} >S_{\text{s, cutoff} }$) with the biased in-situ systems. After sampling, we find that the best cutoff value is $S_{\text{s, cutoff}} = 30$, which produces $p$-values that are all greater than $0.05$ (i.e., $95\%$ confidence). Namely, this criterion suggests that systems with $S_{\rm s} < 30$ ($S_{\rm s} > 30$) are likely to have a formation history dominated by migration (in-situ), which we display in Figure \ref{fig:SMs_ss}. Leveraging this criterion, we find that $62\%$ of our observed systems likely exhibited planetary migration as their dominant formation mechanism.

A similar, but more restricted argument can be made using $S_{\rm c}$ (see Appendix \ref{subsubsec:split_by_sc}).
By splitting the Kepler sample based on these cutoff values, we discover that it is also consistent with the simulated period-ratio distribution as well as other orbital characteristics, which we outline in Section \ref{app: char_of_formation_mechanisms}.

%Based on the empirical analyses above, we further sharpen these constraints to assert that a robust signature of a multi-planetary system's formation history is its orbital spacing, $S_{\rm s}$. Namely, if $S_{\rm s} < 30$ the system was likely formed via planetary migration, and if $S_{\rm s} > 30$ then the system was likely dominated by in-situ assembly.

%\subsubsection{Summary of Predicting Formation Mechanism}\label{subsubsec:summary_of_formation}

%Notably, the $S_{\rm c}$ and $S_{\rm s}$ criteria identified above lie within the predicted ranges from Section \ref{subsec:predictions}. We plot the Kepler systems split by the $S_{\rm c}$ cutoff (top row) and $S_{\rm s}$ cutoff (bottom row) in Figure \ref{fig:SMS_{\rm s}s}. By comparison, we find that the $S_{\rm s}$ cutoff provides consistency between Kepler systems and simulations for both the $S_{\rm s}$ and $S_{\rm c}$ CDFs. Therefore, based on the two above criteria, we conclude that the $S_{\rm s}$ criterion is most robust for predicting the dominant formation mechanism for an observed multi-planetary system. Crucially, we have found that both SMs supply insights into the formation history of observed systems, and can be leveraged to gain insights into the dynamical history of planets within the system. 

By combining the empirical SM ranges above with the theoretical power-law ranges provided in Figure \ref{fig:kep_MGIS_SM_scatter}, we can predict the dominant formation mechanism of nearly all multi-planetary systems in our sample. Leveraging the SMs in this way has shown to be highly robust in gaining insights into the formation mechanism of any observed multi-planetary systems.
%We 

One can infer the dominant formation mechanism using the following systematic process:
\begin{enumerate}
    \item If the calculated ($S_{\rm s}, S_{\rm c}$) values lie in a region of the $S_{\rm s}-S_{\rm c}$ diagram (Figure \ref{fig:kep_MGIS_SM_scatter}) that is unique to one formation mechanism (i.e., in a shaded region without overlap) in accord with the multiplicity of the system, then that is the dominant formation mechanism.
    \item If ($S_{\rm s}, S_{\rm c}$) values lie in a region of the $S_{\rm s}-S_{\rm c}$ diagram that is either occupied by multiple shaded regions with the same multiplicity or unshaded regions, then employ the $S_{\rm s} = 30$ criteria. Namely, if $S_{\rm s} <30$, then the formation of the systems was likely dominated by planetary migration. Conversely, if $S_{\rm s} > 30$, the formation of the systems was likely dominated by in-situ assembly.
\end{enumerate}

%\textbf{We note that, for systems with $N=4$, the in-situ simulations largely overlap with other regions, therefore introducing ambiguity into that region of the $S_{\rm s}-S_{\rm c}$ plot. Furthermore, most of such high-multiplicity systems have $S_{\rm s} < 30$, which make the second criteria less effective for such systems. Lastly, it is crucial to understand that these criteria apply to the observed planetary systems, accounting for observational biases inherent to Kepler Space Telescope Observations. This point is crucial for the in-situ systems because their biased observations often contain 2-4 times fewer planets than the inherent system (e.g., see Figure \ref{fig:bias_arch}. Specifically, most ($97\%$) of the in-situ systems without bias would be classified as having a formation mechanism consistent with migration, based on their $S_{\rm s}$ value. Nevertheless, the current transiting technologies do not have the capabilities of observing such distant and small-sized planets. Therefore, our focus on the biased architecture provides a strong prediction to uncover the formation history of planetary systems as they are observed.  } 

Identifying a formation mechanism of planetary systems with high multiplicity only by $S_{\rm s}$ and/or $S_{\rm c}$ could be hard. In the current analysis, such identification becomes possible largely for planetary systems that exhibit 2-3 planets after biasing (Figure \ref{fig:kep_MGIS_SM_scatter}). As described in Section \ref{subsec:biasing_sim_data}, biasing affects planetary systems formed by in-situ considerably, while it does not for systems formed by migration. This occurs because the former systems tend to have planets that are less massive (i.e., smaller) with non-zero inclinations and/or long orbital periods. These planets are susceptible to observational biases and have a high chance of not being detected by transit observations. The resulting change in orbital architecture is still significant enough to trace the formation history, as shown in this work.

For high-multiplicity planetary systems, however, if most of the planets in the systems are observed, both the formation mechanisms lead to tightly packed systems; observed planets should be reasonably massive, and their inclinations/semimajor axes should be small.
This configuration makes it difficult to specify the formation mechanism only by $S_{\rm s}$ and/or $S_{\rm c}$.
In fact, for systems with $N=4$, the in-situ simulations largely overlap with other regions (Figure \ref{fig:kep_MGIS_SM_scatter}), therefore introducing ambiguity into that region of the $S_{\rm s}-S_{\rm c}$ plot. Furthermore, most of such high-multiplicity systems have $S_{\rm s} < 30$, which make the second criteria less effective for such systems. 
Also, we find that most ($97\%$) of the in-situ systems without biases would be classified as having a formation mechanism consistent with migration, only based on their $S_{\rm s}$ value.
Different quantities (e.g., bulk and atmospheric compositions) are needed to reliably identify a formation mechanism of planetary systems with high multiplicity.
The current transiting technologies do not have the capability to observe small-sized planets with high inclination and/or large orbital periods. Therefore, our focus on the biased architecture provides a strong prediction to uncover the formation history of planetary systems observed currently.

\subsection{Application to Planetary Systems in Habitable Zones}\label{subsec:application_to_systems}
In the previous section, we established methods to gain insight into planetary formation history. Here, we apply these criteria to multi-planetary systems that have at least one planet in the habitable zone (hereafter `habitable systems') to determine their dominant formation mechanisms. The habitability criteria we employ are conservative and seek planets that are likely to exhibit a rocky composition and support surface liquid water \citep[habitable zones determined by][for M-dwarfs and other main-sequence stars, respectively]{Kopparapu13MD, Kopparapu13}. For our systems\footnote{For the search, we used The Habitable Worlds Catalog:
https://phl.upr.edu/hwc}, we look for planets with $0.5 < R_{\rm p} / R_\oplus \leq 1.6$  or $0.1 < M_{\rm p,min} / M_\oplus \leq 3$, or slightly more optimistically, $1.6 < R_{\rm p}/R_\oplus \leq 2.5$ or $3 < M_{\rm p,min}/M_\oplus \leq 10$. We further detail the identified systems and information about their habitability in Appendix \ref{app:hab_systems}.

We calculate the SMs ($S_{\rm s}, S_{\rm c}$) of $13$ habitable systems identified from the above criteria. For these planets, if there was no estimate provided for the planet mass ($M_{\rm p}$), we determine the mass in two ways, depending on the method of detection. If the planet was detected using radial velocity methods and has a reported $M_{\rm p}\sin i$ value, we average this over an inclination distribution that is uniform in $\cos i$ to arrive at a median $M_{\rm p}$ estimate. On average, this leads to a mass estimate that is $16\%$ larger than the reported $M_{\rm p}\sin i$ \citep[e.g.,][]{Zechmeister19}. If the planet was detected using any other method, we apply the \citet{CK17} mass-radius relation to estimate the $M_{\rm p}$ for planets that only have a reported radius.
This follows naturally because the \citet{CK17} mass-radius relation was employed to derive radius estimates for the simulations. We often apply this method for planets detected by transit. The statistical measures of the 13 systems, along with their predicted formation mechanisms, are listed in Table \ref{table:hab_SMs}.

%For the habitable systems in Table \ref{table:hab_SMs} (and more generally for all systems),

Through the criteria discussed in the previous section, we predict that systems $GJ~667C$ ($1$), $GJ~1061$ ($6$), $GJ~273$ ($9$), $LP~890~9$ ($12$), 
and $LHS~1140$ ($13$) were formed via in-situ assembly. From the second criterion, we infer that systems $K2-72$ ($8$) and $TOI-700$ ($10$), which exhibit $S_{\rm s} < 30$, were dominated by planetary migration.

We note that if the multiplicity is not covered by the simulated sample, as is the case for systems $Trappist~1$ ($2$), $Kepler-186$ ($5$), and $Kepler-62$ ($11$), then we cannot make a strong prediction regarding the formation because such systems are not represented in our biased simulation data. 

$Proxima~Centauri$ ($7$) lies in a region of Figure \ref{fig:kep_MGIS_SM_scatter} that is unoccupied by our simulations, so we leave the formation mechanism ambiguous. $Teegarden’s~Star$ ($3$) and  $GJ~1002$ ($4$) lie in regions of Figure \ref{fig:kep_MGIS_SM_scatter} that are occupied by both migration and in-situ systems, so we cannot make any definite predictions about their formation, though we suggest that the former may have experienced early migration followed in-situ assembly, as described in \citet{Hansen12}. Moreover, modern telescope technology strongly favors the detection of close-in planets and is not currently powerful enough to identify Earth-like habitable planets out to $1$~au. Therefore, the applications of our method to determine the formation mechanism of habitable planets are restricted to observable planets that are relatively close-in, similar to the population described above.

% \begin{figure*}
% \includegraphics[width=1.0
% \textwidth]{KNN_feat_import.pdf}

% \figcaption{
% SMs in compared to the high mass and low mass systems.
% \label{fig:feat_import} }

% \end{figure*}

% \subsection{Comparing to KeplerKepler Observations}\label{subsec:biasing_sim_data}
% This is the next step.

% \section{Discussion \& Conclusions}\label{sec:conclusions}
% Discuss and Conlcude, include this after comparing to kepler.

%Need a concluding sentence?

\begin{deluxetable}{lcccc}
%\begin{table}
\tablecaption{Statistical Measures of Habitable Systems\label{table:hab_SMs}}
\tablewidth{0pt}  
\tablehead{
\colhead{Planetary} &  
\colhead{$S_{\rm s}$} &  
\colhead{$S_{\rm c}$} & 
\colhead{Predicted}  &
\colhead{Multiplicity}  
\\
\colhead{System} &  
\colhead{} &  
\colhead{} &
\colhead{Formation} &
\colhead{}
}
\startdata
1. $GJ~667C$  & $44.1$  & $26.4$ &  in-situ & $2$\\
2. $Trappist$--$1$  & $18.6$  & $15.8$ & -& $7$\\
3. $Teegarden's~Star$ & $43.0$ & $66.7$ &  - & $2$\\
4. $GJ~1002$  & $34.4$ & $93.6$ & in-situ/migration& $2$\\
5. Kepler--$186$  & $28.3$ & $9.2$ & - & $5$\\
6. $GJ~1061$  & $30.2$ & $37.8$ &  in-situ  & $3$\\
7. $Proxima~Centauri$  & $43.4$ & $124.2$ & -  & $2$\\
8. $K2$--$72$  & $20.8$ & $35.6$ & migration  & $4$\\
9. $GJ~273$  & $54.5$ & $31.6$ &  in-situ  & $2$\\
10. $TOI$--$700$  & $16.9$ & $103.0$ & migration  & $4$\\
11. Kepler--$62$  & $26.5$ & $6.5$ & -  & $5$\\
12. $LP~890$--$9$  & $46.1$ & $37.4$ & in-situ  & $2$\\
13. $LHS~1140$  & $62.4$ & $19.9$ &  in-situ  & $2$\\
\enddata
\end{deluxetable}

\section{Conclusions and Summary} \label{sec:conclusions}
In this study, we analyze two sets of planet-formation simulations to identify unique characteristics within each one. The first set assesses planetary migration while the other tests in-situ assembly as the dominant formation mechanism.  We introduce biases to model artificial detections akin to observations from the Kepler Space Telescope. For each biased system, we calculate the Statistical Measures (SMs), which include mass concentration ($S_{\rm c}$) and orbital spacing ($S_{\rm s}$). After biasing the simulations, we compare them to a catalog of over $250$ Kepler systems, aiming to predict the dominant formation mechanism of these observed systems, with a particular focus on systems with habitable planets. We summarize our analysis and findings as follows:
\begin{enumerate}
    \item Through theoretical calculations, %we provide analytic predictions for the ranges of SMs that we predict will be unique to the different formation mechanisms. Namely, 
    we predict that if a system has a formation history dominated by planetary migration, it will exhibit $S_{\rm c} \ga 60$ and $S_{\rm s} \la 40$, and if it was dominated by in-situ assembly, it will exhibit $S_{\rm c} \la 300$ and $S_{\rm s} \ga 20$. We further expand on the two-planet case to derive a relationship between $S_{\rm s}$ and $S_{\rm c}$, finding that, to first order, $S_{\rm c} \sim S_{\rm s}^{-2}$, assuming that the planets are in a tight orbit %tight 
    and have similar masses. 
    %Motivated by this result, we later investigate all of our simulations, regardless of spacing and multiplicity, through the lens of a power-law relationship. 
    
    \item %To accurately compare our simulations to observations from the Kepler Space Telescope, we first biased our simulations by taking simulated detections of each with Monte Carlo methods. %We outline a method for applying observational biases by using the \texttt{EPOS} software on the simulated multi-planetary systems. Through Monte Carlo methods, this process reduces a planetary system to only the combination of planets that are detectable by the Kepler Space Telescope. From this, we discuss how biasing changes the planetary architecture of a system. 
    We calculate and compare the SMs for both the intrinsic and biased systems and find that they can differ significantly. Both the $S_{\rm c}$ and $S_{\rm s}$ generally increase %and $S_{\rm s}$ generally decreases 
    after biasing. The architecture before and after is displayed in Figure \ref{fig:bias_arch}, and the SMs before and after are displayed in Figure \ref{fig:SMs}.
    
    \item %After biasing our simulations, we analyze their statistical measures to identify patterns between the different formation mechanisms. In the left panel of Figure \ref{fig:kep_MGIS_{\rm s}M_scatter}, 
   % we 
    We create a scatter plot of the SMs of simulated systems
    %. From the apparent linearity in the plot, which is in log scale, we support our previous prediction 
    and find that they obey power law relations, $S_{\rm c} \sim S_{\rm s}^\beta$ (the left panel of Figure \ref{fig:kep_MGIS_SM_scatter}).
    %Specifically, the different formation mechanisms and the various multiplicities are all approximately linear in log-log scale, with different values of $\beta$. 
    By performing a least-squares fit, we find that $\beta = -2.03, -2.20, -2.51$ for migration systems with multiplicities of $N=2, 3, 4$, respectively. For the simulated in-situ simulations, we find that $\beta = -2.42, -3.11, -3,93$ for systems with multiplicities of $N=2, 3, 4$, respectively. Notably, the $N=2$ migration systems exhibit $\beta \approx 2$, which is consistent with the assumptions and analytic predictions provided earlier.

    \item %Guided by the ranges of $S_{\rm c}$ and $S_{\rm s}$ provided in Section \ref{subsec:predictions}, we 
    We attempt to split our entire Kepler sample of multi-planetary systems by a criterion that depends on either $S_{\rm c}$ or $S_{\rm s}$. 
    %For each SM, we test various values to split the sample and select the value that maximizes the p-value from the KS test. As a result, we 
    We find that separating the Kepler sample about $S_{\rm s} = 30$ is optimal to discriminate between formation mechanisms. %Of the two, we find that $S_{\rm s}$ provides a more robust signature. 
    From Figure \ref{fig:SMs_ss},
    %we find that observed systems with $S_{\rm s} <30$ have both $S_{\rm c}$ and $S_{\rm s}$ distributions consistent with our simulations. Therefore, 
    we conclude that systems with $S_{\rm s}<30$ ($S_{\rm s}>30$) are likely to have a formation history dominated by migration (in-situ). 
    
    % Similarly, we also find that systems with $S_{\rm c}>70$ ($S_{\rm c}<70$) are likely to have a formation history dominated by migration (in-situ). Both criteria are consistent with the ranges predicted analytically.%, and serve as an extra condition to determine the dominant formation mechanism for a multi-planetary system.
    
    \item Finally we leverage our analysis of the SMs to predict the dominant formation mechanism of $257$ observed multi-planetary systems, including 13 with one or more planets in the habitable zone. Based on the $S_{\rm s}$ criteria, we conclude that $62\%$ of our observed sample experienced planetary migration.
    Based on Figure \ref{fig:kep_MGIS_SM_scatter} and Table \ref{table:hab_SMs},  
    we also find that most of our habitable systems formed via in-situ assembly. This could suggest that the dynamical history that comes with in-situ assembly %increases 
    might increase the formation and detection rates %rate 
    of terrestrial planets in the habitable zone.
\end{enumerate}
    
In this work, we demonstrate the profound capabilities of the statistical measures to predict the formation history of multi-planetary systems. %By calculating the mass concentration ($S_{\rm c}$) and orbital spacing ($S_{\rm s}$) of the present-day, observed architecture of a planetary system, one can apply our methods to gain insight into the dominant mechanism that pushed the formation of the system.
We also find that the SMs of the biased systems differ from those of the intrinsic system (e.g. Figure \ref{fig:bias_arch}). If these changes are moderately predictable, one can disentangle the characteristics of the intrinsic systems based on their observed architecture. We do not pursue these methods of `debiasing' in this work but note that it could be possible for future work.

\section{acknowledgements} \label{acknowledgements}
We thank the anonymous referee for a constructive report. C.S. thanks the SURF@JPL program. This research was carried out at the Jet Propulsion Laboratory, California Institute of Technology, under a contract with the National Aeronautics and Space Administration (80NM0018D0004). Y.H. is supported by JPL/Caltech.

\clearpage
% APPENDIX STARTS HERE

\appendix
\section{A simplified expression of mass concentration for the two-planet case}\label{app:SMs_2planets}

We here briefly describe how a simplified expression of mass concentration can be derived for the two-planet case.

Mass concentration for the two-planet case is written as (equation (\ref{eq:Sc}))
\begin{equation}
    \label{eq:Sc_2planets}
    S_{\rm c} = \mbox{max} \left( \frac{ M_{{\rm p},1} + M_{{\rm p},2} }{ M_{{\rm p},1} [ \log_{10}{(a/a_{{\rm p},1})}]^2 + M_{{\rm p},2} [ \log_{10}{(a/a_{{\rm p},2})}]^2 }  \right).
\end{equation}
A maximum value can be achieved when the denominator takes a minimum value.
Taking the derivative of the denominator, we find that it becomes possible when 
\begin{equation}
  \log_{10}a = \frac{ M_{{\rm p},1} \log_{10} (a_{{\rm p}, 1}) + M_{{\rm p},2} \log_{10} (a_{{\rm p},2}) }{ M_{{\rm p},1} + M_{{\rm p},2} }. 
 \end{equation}
Plugging the above $\log_{10}a$ into equation (\ref{eq:Sc_2planets}), 
then mass concentration is expressed as
\begin{equation}
    S_{\rm c} =  \frac{ ( M_{{\rm p},1} + M_{{\rm p},2} )^2 }{ M_{{\rm p},1} M_{{\rm p},2} }  \frac{1}{[ \log_{10}{(a_{{\rm p},2}/a_{{\rm p},1})}]^2}.  
\end{equation}

\section{Additional discussions about the outcome of $\EPOS$}\label{app:EPOS}

We provide a brief additional examination of the outcome of $\EPOS$.

After biasing by $\EPOS$, the multi-planetary systems are often reduced to the systems that have planets with larger masses and shorter periods.

\begin{figure*}
\centering
\includegraphics[width=0.5
\textwidth]{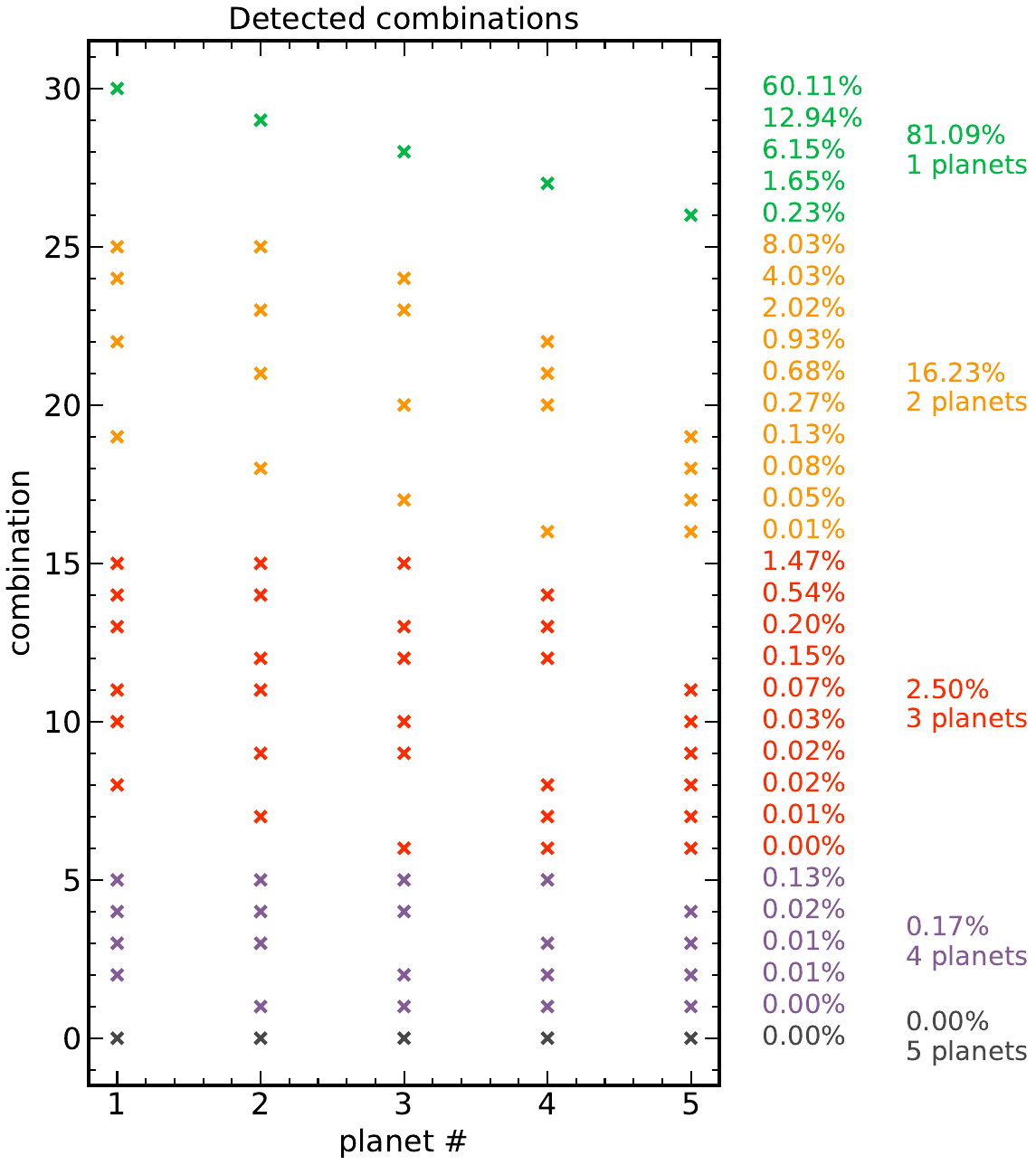}

\caption{Detection combinations and probabilities for an example multi-planetary system. Observing the innermost planet is the most likely scenario ($60.11\%$). Observing the inner two planets is the most probable ($8\%$) multi-planet observation to be made by Kepler. }.\label{fig:detection_eg} 
\end{figure*}

In Figure \ref{fig:detection_eg}, we display the probabilities of different combinations of planets being observed by Kepler. In this example, it is most likely ($81.09\%$) that only one planet will be recovered, 
with the highest probability of $\sim60\%$ that only the closest planet will be observed. However, there is a $16.23\%$ chance that two planets will be recovered, with the most visible combination being the inner two planets (planets $1~\&~2$). The probability is highly dependent on the orientation of the system \citep{EPOSI}. We also find that the shorter-period planets, regardless of the planet's mass, 
have higher probabilities of being detected.
Moreover, the planets with relatively low mutual inclinations ($i\lsim 5^\circ$) are more easily observed with simulated Kepler transit detections.

\section{Classification of exoplanetary systems} \label{app:class}

We provide additional discussions about how observed exoplanetary systems can be classified so that their formation origins are identified.

\subsection{Splitting by $S_{\rm c}$}\label{subsubsec:split_by_sc}

\begin{figure*}
\includegraphics[width=0.95
\textwidth]{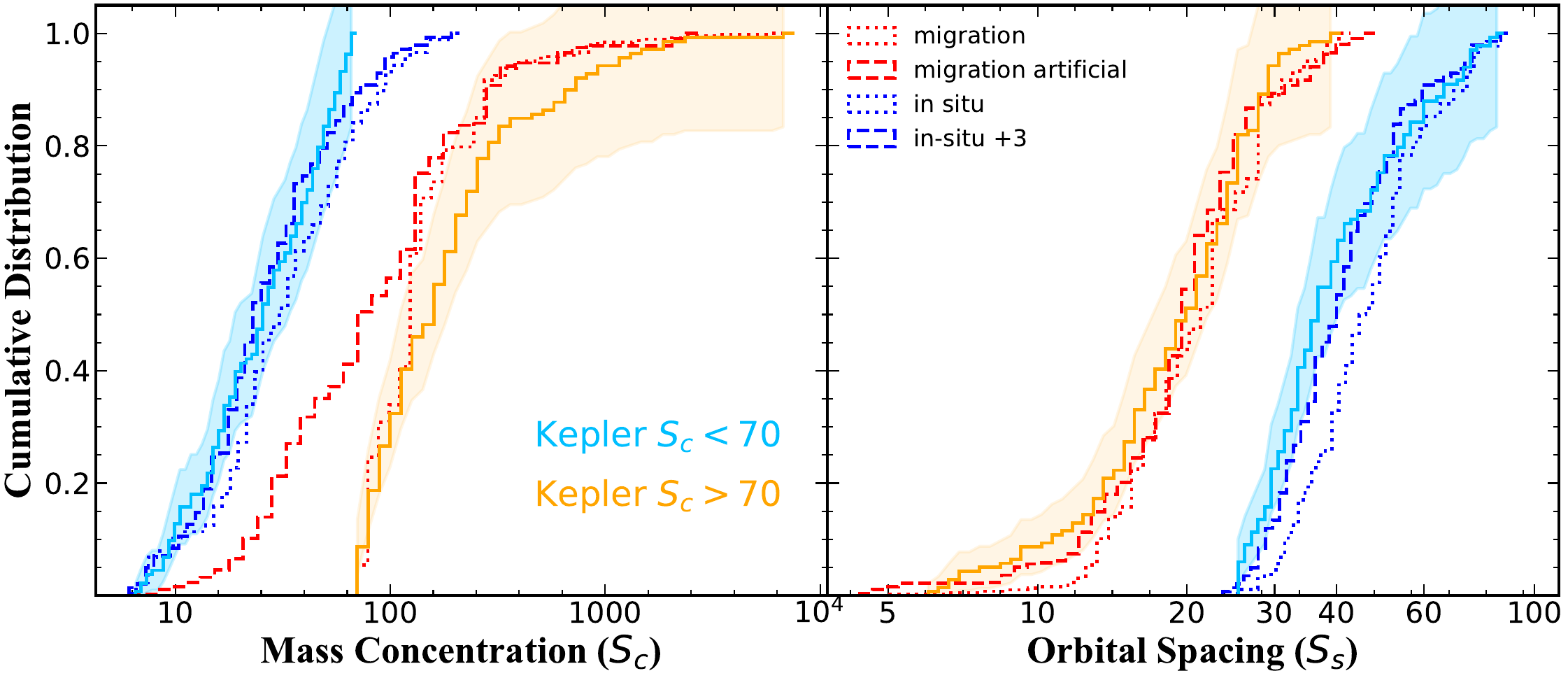}
\figcaption{ 
The cumulative distributions of the mass concentration ($S_{\rm c}$; left column) and orbital spacing ($S_{\rm s}$; right column) of the biased simulation data as well as
the observed Kepler sample, which is split into two subgroups. 
The curves are the same as Figure \ref{fig:SMs_ss}, but here, we split the Kepler by those with $S_{\rm c} > 70$ (orange) and those with $S_{\rm c}<70$ (light blue).}
\label{fig:SMs_sc}

\end{figure*}
By inspection of Figure \ref{fig:SMs}, we find that the minimum mass concentration of the two-planet migration systems is $S_{\rm c} \sim 70$. Therefore, we first filter the Kepler multi-planetary systems solely under the condition that $S_{\rm c} \geq 70$. The total of $301$ planets in $139$ systems satisfy this criterion, making them potentially more consistent with migration-dominated formation. For the remaining systems with $S_{\rm c} < 70$, we further constrain them to have $S_{\rm s}$ bounds that are matched with the limitations of the in-situ simulations. The total of $309$ planets in $133$ systems satisfy these criteria, making them potentially more consistent with an in-situ-dominated formation scenario. 

In  Figure \ref{fig:SMs_sc} we plot both of the aforementioned subgroups derived from our Kepler sample alongside the biased theoretical systems. We first plot the curve of the SMs for the in-situ (dashed blue) and migration (dashed red) after they have been biased (see Section \ref{subsec:biasing_sim_data} for details on the biasing). Next, in light blue, we plot the Kepler systems with $S_{\rm c} < 70$ and $S_{\rm s}$ within the range of the in-situ data. In orange, we show the Kepler systems with $S_{\rm c} \geq 70$. The shaded regions for both of these curves represent the Poissonian error ($\sqrt{N}$) for each bin. 

We find that Kepler systems with $S_{\rm c} \geq 70$ have both $S_{\rm c}$ and $S_{\rm s}$ distributions that are broadly consistent with those from the migration simulations. Namely, we can leverage this consistency to determine that the observed multi-planetary systems in our Kepler sample with $S_{\rm c} \geq 70$ reflect a formation history dominated by planetary migration. Moreover, the remaining Kepler systems (light blue in Figure \ref{fig:SMs_sc}) have an $S_{\rm c}$ distribution consistent with the simulated in-situ systems. %\sout{Specifically, the $S_{\rm c}$ curves exhibit nearly} \sout {identical progression, whereas the $S_{\rm s}$ curves have the same slope for orbital spacings greater than $\sim 50$.}

We find that the in-situ systems with higher multiplicities occupy the region of abundance $S_{\rm s} \in \{25, 50\}$. This could suggest that the Kepler systems with higher multiplicities, which are missing from the sample, would also occupy this region. We test this hypothesis to find that, when we invoke a larger number of higher multiplicity ($\geq 3)$ systems into our biased in-situ sample, the curves overlap nearly identically. The dashed green curves in Figure \ref{fig:SMs_sc} represent the in-situ simulations sampled so that their proportion of 3-planet systems matches the ratio in our observed sample.

%From the stark similarities between the observed and simulated samples, we infer that the mass concentration and orbital spacing of observed systems are strong signatures of a multi-planetary systems' formation history. We now refer to the systems in our Kepler sample with $S_{\rm c} < 70$ as `Kepler In-Situ (IS)' and the subgroup with $S_{\rm c} < 70$ as `Kepler Migration (MG)'. 

Although this $S_{\rm c}$ cutoff was initially determined from the minimum $S_{\rm c}$ of the migration simulations, we find that it is statistically optimized. Namely, we perform the KS test for the cumulative distributions of $S_{\rm c}$ and $S_{\rm s}$ compared to the distributions of the observed Kepler systems to determine which $S_{\rm c}$ cutoff between $S_{\rm c}\in \{50,90\}$. From this, $S_{\rm c}=70$ gives the largest KS test $p$ values. The $S_{\rm c}=70$ cutoff produces $p$-values greater than $0.05$ (assuming $95\%$ confidence), which does not support the hypothesis that the two distributions came from different parent distributions. 

\subsection{Machine Learning Classification}\label{app:other_ML_models}

\begin{figure*}
\includegraphics[width=1.0
\textwidth]{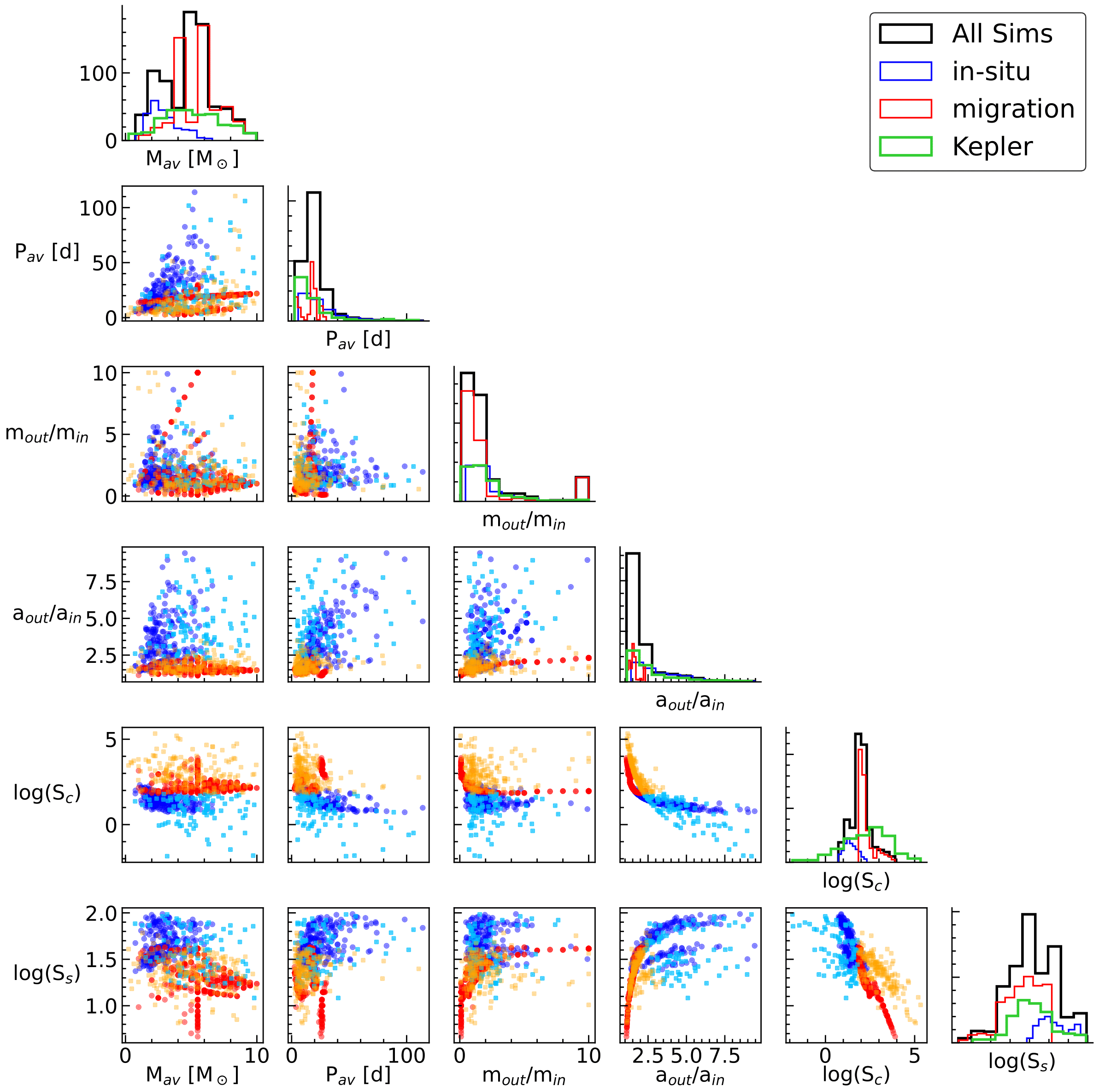}

\figcaption{
Triangle plot of the parameters used to train KNN model to predict the dominant formation mechanism. The features of the multi-planetary systems that were inputted into the various ML models include the (1) average period ($P_{av}$), (2) mass ratio between the farthest and closest planet ($m_{out}/m_{in}$), (3) semi-major axis ratio between the farthest and closest planet ($a_{out}/a_{in}$), (4) the mass concentration ($S_{\rm c}$), and (5) the orbital spacing ($S_{\rm s}$). 
\label{fig:KNN_params} }

\end{figure*}

We verify our classification in Sections \ref{subsec:identifying_formation_mechanism} and \ref{subsubsec:split_by_sc}, using machine learning.
We test various machine learning methods trained on the simulation data and apply them to the Kepler data. These supervised methods are trained on the following features of the multi-planetary system: the mass concentration ($S_{\rm c}$), the orbital spacing ($S_{\rm s}$), the average period ($P_{av}$), the mass ratio ($M_{out}/M_{in}$), the period ratio ($P_{out}/P_{in}$), and the multiplicity. We present a corner plot of these parameters and their correlations with each other in Figure
\ref{fig:KNN_params}. 

After training and testing the model with the simulation data, we find that the accuracy is effectively unchanged if we remove the latter three features. With only the first three features in consideration, we employ several machine learning algorithms and assess their capability to predict the dominant formation mechanism of multis.

\subsubsection{k-Nearest Neighbors}\label{subsec:kNN}

% \subsubsection{k-Nearest Neighbors Classifier}\label{subsubsec:kNN}

The first machine learning model we use is the k-Nearest Neighbors (kNN) algorithm. This model as a method of classifying detections of exoplanets has been deemed highly fruitful in previous work \citep{KNN_paper}. After preprocessing, training, and testing our data with the kNN model, we reach an accuracy of $96\%$. We discover that mass concentration is one 
crucial predictor of the formation mechanism of the system. However, the fact that the average period is highly correlated with the formation mechanism might be a result of the bias in the migration simulations. Namely, these simulations exhibit smaller periods by design, so we can ignore this feature as well. Interestingly, a majority ($\sim65\%$) systems are classified as having a formation process dominated by migration of the inner planets. Most importantly, the kNN model also supports that the mass concentration and orbital spacing are the two most important features %, of those tested, 
in predicting the dominant formation mechanism.

\subsubsection{Neural Network Model}\label{subsubsec:NN}
As an alternative to the already highly accurate kNN model, we also test a manually designed neural network (NN) for the binary classification process. The NN is made with only one hidden layer containing 6464 nodes to avoid overfitting, especially considering our small number of features and relatively low sample size. After training the model for 20 epochs, we reach an accuracy of 92.8.%92.8\%, which is comparable to that of the kNN model. This model also only uses the three features outlined and also provides a feature importance scale. As the epochs progress during the model training, both the loss and accuracy curves asymptote for the latter half of the epochs. The leveling out of the loss and accuracy over time is an indicator of the model recognizing important patterns, instead of simply learning the data. The bottom half of this plot shows the relative importance of each input feature that the model uses to predict the formation mechanism for each trial. We find consistency with the results of our kNN model, in that the mass concentration is still the strongest feature, with the orbital spacing and average period being below. However, in the NN model, the orbital spacing is weighted more heavily, and more important in predicting whether a system is in-situ or migration-dominated. \\

% \begin{figure}
% \includegraphics[width=0.5
% \textwidth]{NN_trends.pdf}
% \includegraphics[width=0.5
% \textwidth]{NN_shap_values.png}

% \figcaption{
% The output of training the Neural Network that was designed for binary classification.
% \label{fig:NN_plots} }

% \end{figure}

\subsubsection{Other ML Classifiers}\label{subsubsec:other_ML}
To be complete, we also test Support Vector Machine (SVM) and logistic regression models to see if any of them are more accurate than our other algorithms. With optimization, we find that these models are unable to reproduce accuracies as high as the previous models, so we do not apply them to the Kepler data.

% \begin{figure}
% \centering
% \includegraphics[width=0.8
% \textwidth]{KNN_feat_import.pdf}

% \figcaption{
% SMs in compared to the high mass and low mass systems.
% \label{fig:feat_import} }

% \end{figure}

\section{Characteristics of Different Formation Mechanisms}\label{app: char_of_formation_mechanisms}

\begin{figure*}
\centering
\includegraphics[width=0.60
\textwidth]{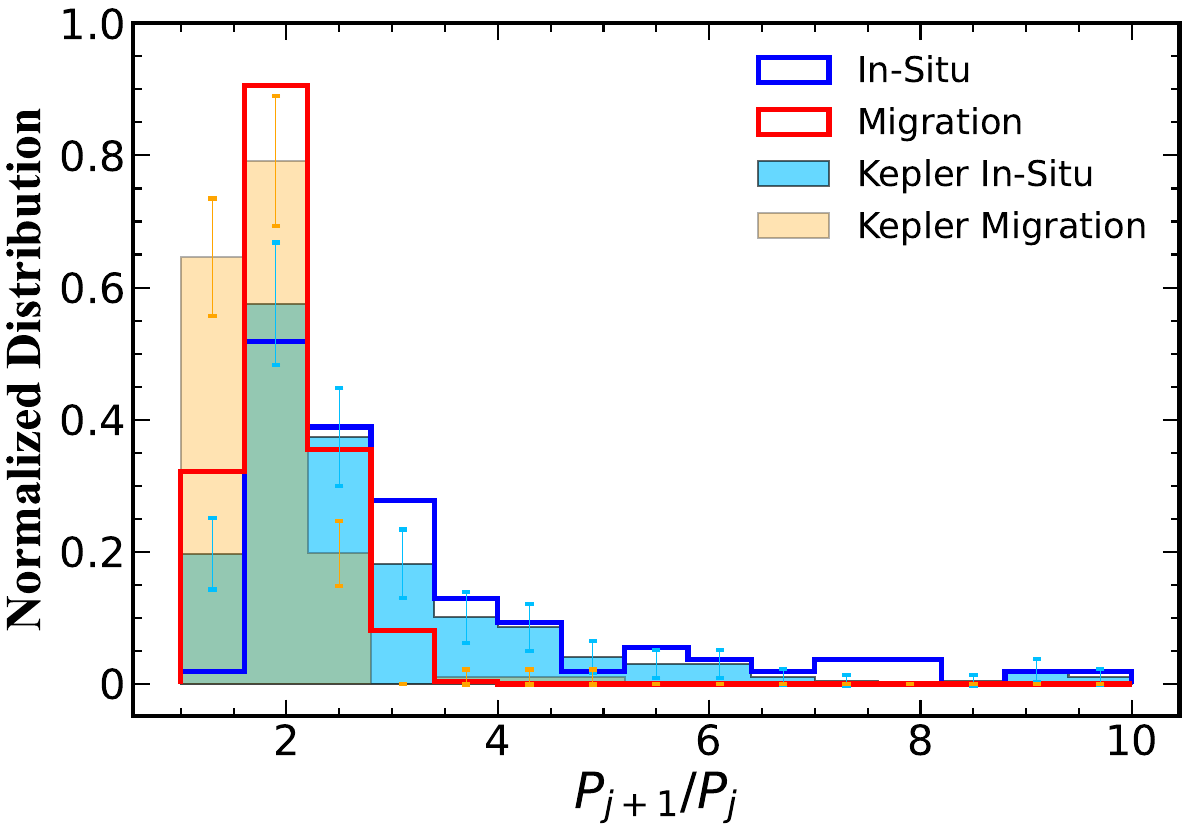}
\figcaption{
The normalized density distribution of the period ratios ($P_{j+1}/P_j, j\in \mathbb{Z}$) for all adjacent planets. The migration (solid red) and in-situ (solid blue) simulations are compared to the Kepler migration systems (light orange) and Kepler in-situ systems (light blue). The observed systems have Poisson error bars ($\sqrt{n}$ for each bin) overplotted.
\label{fig:kep_MGIS_pratio} }

\end{figure*}

Planets that migrate inward are likely to exhibit long resonant chain structures where the period ratios of neighboring planets are simply integer ratios \citep[e.g.,][]{Terquem07, Raymond08, Izidoro17, Ormel17, Kajtazi23}. We look for planets close to such mean motion resonances %during their migration 
within the Kepler systems. We identify a pair of adjacent planets as having a resonant structure if their period ratio ($P_{j+1}/P_j$) is within $10\%$ of the 2:1, 4:3, or 3:2 resonant states. See Figure \ref{fig:kep_MGIS_pratio} for the distributions of period ratios for all of our models and observations.
% We plot the period ratios for all planets in the Kepler IS and Kepler MG systems and compare them to the results of the migration and in-situ simulations (left panel of Figure \ref{fig:kep_MGIS_pratio_mult}.

%Planet formation theory suggests that planets exhibiting migration are more likely to be captured by orbital resonances \citep{Peale76, Goldreich80}. In accord, we 
We find that the Kepler migration systems -- with radial mass concentrations larger than $70$ --  exhibit more systems with resonant structures than the Kepler in-situ systems. Similar to the migration simulations, the Kepler migration planets have peaked period ratios near the resonant values of $2/1$, $4/3$, and $3/2$. Specifically, we find that $73\%$ ($118/162$) of the Kepler migration planet-pairs exhibit one of these primary mean motion resonances, whereas only $15\%$ ($26/176$) of the Kepler in-situ systems have planetary pairs with these integer period ratios. 

This feature appearing in orbital characteristics can be viewed as an additional confirmation that our classification of observed exoplanetary systems by the SMs is reasonable.

%The number of planets in a system (i.e., the planet multiplicity) also conveys information about the formation and evolution of multi-planetary systems. The dynamical interactions amongst the various planets may imprint rich signatures in the present-day architecture of the system \citep[e.g.,][see latter for a detailed review]{Chatterjee08, Juric08, Pu15, Zhu21}.
% We compare the multiplicities of the different Kepler groups in Figure \ref{fig:kep_MGIS_pratio_mult}. Interestingly, $82\%$ ($122/139$) of the Kepler MG systems have a multiplicity of 2. For the Kepler IS systems, only $74\%$ ($98/133$) have a multiplicity of 2, which is very close to the $77\%$ for the biased in-situ systems. Therefore, the Kepler IS systems, because they are consistent with our in-situ sample, are likely to have an intrinsic multiplicity similar to the in-situ simulations. 
%Kepler IS systems may exhibit lower mass concentrations due to interactions with hidden planets. This can reveal that the true multiplicity for these systems is larger \citep[i.e., between 4-10 planets][]{Hansen12}. On the contrary, our results indicate that an observed system with only two close-in, massive planets makes a system more likely to have a formation history dominated by two-planet migration, without much effect from hidden planets.

\section{Habitable Systems}\label{app:hab_systems}
\begin{enumerate}
    \item $GJ~667C$ is proposed to host six (or even seven) planets with varying masses and orbital periods \citep{Anglada12, Delfosse13}. Doppler measurements and fitting analyses identify three super-Earth systems in the habitable zone, with highly constrained planetary parameters \citep{Anglada13}. The SMs ($S_{\rm s}$, $S_{\rm c}$) for this system are ($41.9$, $26.4$).
    \item $Trappist$--$1$ is a system of seven temperate terrestrial planets with the six inner planets forming a near-resonant chain \citep{Gillon17}. From the resonant architecture, the formation is likely to be more consistent with a migration scenario, as outlined in \citet{Gillon17}. However, the migration simulations that we employ are restricted to only two-planet systems, which makes this system a poor fit to compare to our sample. Regardless, we find that the orbital spacing of the Trappist~1 system is in the range of the migration simulations, and has no overlap with the in-situ simulations. The SMs ($S_{\rm s}$, $S_{\rm c}$) for this system are ($18.6$, $15.8$).
    \item $Teegarden's~Star$ is a star in the nearby solar neighborhood with evidence from over $200$ radial-velocity measurement for two earth-mass planet candidates. The CARMENES search for exoplanets identified these planets, which have orbital periods of $4.91$ and $11.4$ d, both with a minimum mass of $1.1$~M$_\oplus$ \citep{Zechmeister19}. One planet is in the conservative habitable zone, with both being in the optimistic habitable zone. The SMs ($S_{\rm s}$, $S_{\rm c}$) for this system are ($43.0$, $66.7$).
    % (SsS_{\rm s}, ScS_{\rm c}) = (43.043.0, 66.766.7)
    \item $GJ~1002$ was reported to host two temperate Earth-size planets in its habitable zone through Radial Velocity detections. The planet, GJ 1002 b (GJ 1002 c) has an orbital period of $10$~d ($21$~d) days, with minimum masses of $1.08$~M$_\oplus$ ($1.38$~M$_\oplus$) \citep{Suarez23}. The SMs ($S_{\rm s}$, $S_{\rm c}$) for this system are ($32.7$, $93.6$). 
    % (SsS_{\rm s}, ScS_{\rm c}) = (32.732.7, 93.693.6)
    \item Kepler--$186$ is a main-sequence star, which initially had four planets detected with radii less than $1.5$~R$_\oplus$ and orbital periods ranging from $3.9$-$22.4$ \citep{Rowe14, Lissauer14}. A fifth planet, Kepler-186f, was then detected in the conservative habitable zone of the host star \citep{Tenenbaum13, Kopparapu13, Quintana14} with an estimated intermediate mass of $1.44$~M$_\oplus$ and period of $130$~days \citep{Quintana14}. The SMs ($S_{\rm s}$, $S_{\rm c}$) for this system are ($28.3$, $9.2$). 
    % (SsS_{\rm s}, ScS_{\rm c}) = (28.328.3, 9.29.2)
    \item $GJ~1061$ is a low-mass star with three dynamically stable planet candidates, with the potential for a fourth hidden planet, though the fourth signal may also be due to stellar rotation \citep{Dreizler20}. The third planet, GJ 1061d lies within the habitable zone of the star and exhibits an equilibrium temperature similar to that of Earth. The planets have periods $3.2$, $6.7$, and $13.0$ days and minimum masses of $1.4$, $1.8$, and $1.7$~M$_\oplus$ \citep{Dreizler20}. The SMs ($S_{\rm s}$, $S_{\rm c}$) for this system are ($28.7$, $37.8$). 
    % (SsS_{\rm s}, ScS_{\rm c}) = (28.728.7, 37.837.8)
    \item $Proxima~Centauri$ is the nearest known star to our solar system and has been proposed to host two or three planets \citep{Anglada16, Li17}, though the presence of the third planet has recently been challenged \citep{Artigau22}. To be conservative, we only consider the first two planets. These two planets have orbital periods of $5.1$ and $11.2$~days and minimum masses of $0.26$ and $1.07$~M$_\oplus$ \citep{Faria22, Suarez20}. The second planet lies in the habitable zone of the star. The SMs ($S_{\rm s}$, $S_{\rm c}$) for this system are ($41.3$, $124.2$). 
    % (SsS_{\rm s}, ScS_{\rm c}) = (41.341.3, 124.2124.2)
    \item $K2$--$72$ is located 67 pc away and is detected to host four planets with sizes similar to that of Earth \citep{Crossfield16, Dressing17}. The third planet in this system, K2-72 c, lies in the conservative habitable zone. 
    The SMs ($S_{\rm s}$, $S_{\rm c}$) for this system are ($35.6$, $19.8$).
    \item $GJ~273$ is the second-nearest known planetary system with two detections, Earth-mass planets detected via radial velocity methods \citep{Astudillo17}. The second of these planets, GJ 273 b, lies in the conservative habitable zone.
    The SMs ($S_{\rm s}$, $S_{\rm c}$) for this system are ($31.6$, $51.8$).
    \item $TOI$--$700$ is an M2.5 dwarf star that hosts three planets discovered by TESS, with four planets \citep{Kirkpatrick91, Gilbert20B, Rodriguez20, Gilbert23}. Two of these planets, TOI-700 d, and TOI-700 e, are Earth-sized planets in the habitable zone of the star \citep{Gilbert20B, Rodriguez20, Gilbert23}. The SMs ($S_{\rm s}$, $S_{\rm c}$) for this system are ($103.0$, $16.0$).
    \item Kepler--$62$ is star that hosts five, earth-sized planets \citep{Borucki13}. Two of the planets, Kepler-62e and -62f, are super-Earth's in the habitable zone that are estimated to have rocky compositions with mostly solid water \citep{Borucki13}. Kepler-62f was confirmed to not be a true detection with updated planetary properties by more recent $Gaia$ results \citep{Borucki19}. The SMs ($S_{\rm s}$, $S_{\rm c}$) for this system are ($6.5$, $26.5$).
    \item $LP~890$--$9$ was detected by TESS \citep{Muirhead18} and later discovered to host two temperate super-Earths via transit detection \citep{Delrez22}. The outer planet, LP 890-9 c, lies in the conservative habitable zone. The SMs ($S_{\rm s}$, $S_{\rm c}$) for this system are ($37.4$, $46.1$).
    \item $LHS~1140$ is a nearby \citep[$10.5$~pc.][]{Gaia18} planetary system with two known small, short-period planets ($LHS~1140~c$, Pc $\sim 3.77$~days) ($LHS~1140~b$, Pb $\sim 24.7$~days). The planets were detected using a combination of transit and radial-velocity methods \citep{Nutzman08, Mayor03}. $LHS~1140~b$ lies within the habitable zone of the star, and has recently-updated planet properties. \citet{LilloBox20} combined the power of ESPRESSO and TESS to further constrain the masses to $6.48 \pm 0.46 M_\oplus$ for $LHS~1140~b$ and $1.78 \pm 0.17 M_\oplus$ for $LHS~1140~c$. Moreover, fitting-analysis methods suggest another planet on a $78.9$~day period and a mass of $4.8 \pm 1.1 M_\oplus$, though it has not been confirmed so we do not include it here. We note that including this third planet does not change the inferred formation mechanism based on our assessments. The SMs ($S_{\rm s}$, $S_{\rm c}$) for this system are ($62.4$, $19.9$).
\end{enumerate}
\clearpage
\bibliography{references}

\end{document}